\documentclass[prl,superscriptaddress,floatfix,twocolumn,showpacs]{revtex4-1}
\usepackage{color}
\usepackage{amsfonts,amssymb,amsmath}
\usepackage[]{graphics,graphicx,epsfig}
\usepackage{amsthm, subfigure}
\usepackage{natbib}
\usepackage[colorlinks]{hyperref}
\hypersetup{
	bookmarksnumbered,
	pdfstartview={FitH},
	citecolor={blue},
	linkcolor={red},
	urlcolor={black},
	pdfpagemode={UseOutlines}
	}

\newcommand{\bra}[1]{\langle #1 \vert}
\newcommand{\ket}[1]{\vert #1 \rangle}
\newcommand{\ketbra}[1]{\vert #1 \rangle\langle #1 \vert}

\newcommand{\sclp}[2]{\langle #1\vert #2 \rangle}

\newcommand{\tr}[0]{\mathrm{Tr}}

\begin{document}
\title{Converting Coherence to Quantum Correlations}

\author{Jiajun Ma}
\email{jiajunma2012@gmail.com}
\affiliation{Center for Quantum Information, Institute for Interdisciplinary Information Sciences, Tsinghua University, 100084 Beijing, China}
\affiliation{Department of Atomic and Laser Physics, Clarendon Laboratory,
University of Oxford, Parks Road, Oxford OX1 3PU, United Kingdom}
\author{Benjamin Yadin}\affiliation{Department of Atomic and Laser Physics, Clarendon Laboratory,
University of Oxford, Parks Road, Oxford OX1 3PU, United Kingdom}
\author{Davide Girolami}
\email{davegirolami@gmail.com}
\affiliation{Department of Atomic and Laser Physics, Clarendon Laboratory,
University of Oxford, Parks Road, Oxford OX1 3PU, United Kingdom}
\author{Vlatko Vedral}\affiliation{Center for Quantum Information, Institute for Interdisciplinary Information Sciences, Tsinghua University, 100084 Beijing, China}
\affiliation{Department of Atomic and Laser Physics, Clarendon Laboratory,
University of Oxford, Parks Road, Oxford OX1 3PU, United Kingdom}
\affiliation{Centre for Quantum Technologies, National University of Singapore, 117543  Singapore}
\affiliation{Department of Physics, National University of Singapore, 2 Science Drive 3, 117551 Singapore}
\author{Mile Gu}
\email{ceptryn@gmail.com}
 \affiliation{School of Physical and Mathematical Sciences, Nanyang Technological University, 21 Nanyang Link, 637371, Singapore}
 \affiliation{Complexity Institute, Nanyang Technological University, 18 Nanyang Drive, 637723, Singapore}
 \affiliation{Centre for Quantum Technologies, National University of Singapore, 117543  Singapore}
 \affiliation{Center for Quantum Information, Institute for Interdisciplinary Information Sciences, Tsinghua University, 100084 Beijing, China}

\begin{abstract}
Recent results in quantum information theory characterise quantum coherence in the context of resource theories.  Here we study the relation between quantum coherence and quantum discord, a kind of quantum correlation which appears even in nonentangled states. We prove that the creation of quantum discord with multipartite incoherent operations is bounded by the amount of quantum coherence consumed in its subsystems during the process. We show how the interplay between quantum coherence consumption and creation of quantum discord works in the preparation of multipartite quantum correlated states and in the model of deterministic quantum computation with one qubit (DQC1).
\end{abstract}

\date{\today}

\pacs{03.65.Ud, 03.65.Ta, 03.67.Ac, 03.67.Mn}

\maketitle
\textit{Introduction} --  Quantum information theory studies the features that make a system inherently quantum. Viewing these features as resources is crucial for developing new quantum technologies. Nonclassical correlations have long been regarded as a key quantum resource. Entanglement was the first such concept to be known about, and later other types of quantum correlations beyond entanglement, notably quantum discord, were discovered. However, notions of quantumness also exist in single systems without referring to correlations. As originally pointed out by Schr\"{o}dinger \cite{cat}, superposition -- nowadays often called quantum coherence -- is a fundamentally quantum property. The rigorous characterization of coherence in the framework of resource theories has been a rather recent development \cite{plenio,gour}, and a subsequent stream of works have identified coherence measures for both theoretical and experimental purposes \cite{us,ger1,noriwit,winter,chita,ger2,pradisc,pracond,luo2,xiao,heng,plenio,me,blind,lqu,mehdi,luo,superreview,gour,newmar,herbut,aberg,luocri,speknat,lost,lost2}.

Resources are often interconvertible: we can trade one for another. Along these lines, can coherence in single systems be traded for quantum correlations? Recently, the relation between coherence and entanglement has been studied in Refs. \cite{ger2,pradisc,us,chita2}. In this work, we investigate the interplay between coherence and quantum discord in multipartite systems. Discord is a recently established kind of quantum correlation \cite{OZ,HV,discrev}, which has generated a great deal of interest and debate \cite{datta2}. Previous results have reported a link between specific coherence and discord-type measures \cite{heng,pradisc,hu}. Here we provide a general relation for consuming coherence in order to build up discord. We prove that for a multipartite system, if the coherence of the global state is a resource which cannot be increased, the cost of creating discord can be expressed in terms of coherence consumption of the subsystems.

We show this mechanism at work in two settings. First, we consider the preparation of a quantum correlated state by applying a sequence of controlled gates to an uncorrelated multipartite coherent state, a standard subroutine for quantum information, computation and metrology  \cite{metrorev,wilde}. Then we focus on the deterministic quantum computation model with one qubit (DQC1) \cite{dqc1}. This model has been widely studied to determine if alternative quantum resources beyond entanglement, e.g., discord, might be employed in quantum computation \cite{eastin,datta,datta1,dakic}. We study the role played by the coherence of the single pure qubit in this model, showing that it is coherence consumption that makes discord production possible. 

\textit{Linking coherence and discord} -- Before stating our results, we review the information-theoretic definitions of coherence and discord. The resource theory of coherence  characterizes  the free resources, i.e., incoherent states and incoherent operations, and the criteria identifying coherence measures \cite{plenio,note}.  The framework has been extended to the multipartite scenario \cite{pradisc,ger2,us,chita}. Given a finite-dimensional system partitioned by $n$ subsystems $\{A_1, A_2, ...,A_n\}$, and a reference product basis $\{\ket{i_{1\ldots n}}:=\ket{i_1}\otimes...\otimes\ket{i_n}\}$,  the incoherent states take the form $\sigma_{A_1...A_n}=\sum_{i_{1\ldots n}}p_{i_{1\ldots n}}\ketbra{i_{1\ldots n}}, \{p_{i_{1\ldots n}}\}\geq0, \sum_{i_{1\ldots n}}p_{i_{1\ldots n}}=1$, and form the set $\mathcal{I}_{A_1...A_n}$. Any other state has non-zero coherence. An incoherent operation is any quantum operation that maps incoherent states to incoherent states. Incoherent states of subsystem $A_k$ are defined with respect to the basis $\{\ket{i_k}\}$ being consistent with the joint basis $\{\ket{i_{1\ldots n}}\}$. For a bipartite system $AB$,  it is possible to study the coherence with respect to a local basis on $A$. We thus define the A-incoherent states with respect to $\{\ket{i}_A\}$  as $\label{partialthe partial incoherent operations_inc}  \sigma_{AB}\in {\cal I}_{B|A}, \sigma_{AB}=\sum_ip_i\ketbra{i}_A\otimes\rho_{B|i}$  \cite{QI}, and call maps $\Lambda_{\mathrm{IC}}^{B|A}$ taking ${\cal I}_{B|A}$ to itself A-incoherent operations.
A measure of coherence $f_C(\rho)$ is a nonnegative function which vanishes for incoherent states and is a nonincreasing monotone under incoherent operations, $f_C(\rho)\geq f_C(\Lambda_{\mathrm{IC}}(\rho))$. This ensures that the resource (coherence) cannot be increased through free (incoherent) operations. A notable class of coherence measures is given by the pseudo-distance of a state to the incoherent (or A-incoherent) state set ${\cal I}$: $C^\delta(\rho)=\min\limits_{\sigma\in {\cal I}}\delta(\rho,\sigma)$, where $\delta$ is a contractive pseudo-distance \cite{geometry}. In particular, we recall the relative entropy of coherence $C(\rho)=\min\limits_{\sigma\in {\cal I}}S(\rho||\sigma)=S(\rho||\Phi^i(\rho))$, where $S(\rho||\sigma)=\tr{}[\rho\log_2\rho]-\tr{}[\rho\log_2\sigma]$ is the quantum relative entropy and $\Phi^i(\rho)=\sum_i \ket{i}\bra{i}\rho \ket{i}\bra{i}$ is the dephased state in reference basis $\{\ket i\}$. For the A-incoherent case we have  $C_{B|A}(\rho_{AB})=\min\limits_{\sigma_{AB}\in \mathcal{I}_{B|A}}S(\rho_{AB}||\sigma_{AB})=S(\rho_{AB}||\Phi^i_A(\rho_{AB}))$,  where $\Phi^i_A(\rho_{AB})=\sum_ip_i\ketbra{i}_A\otimes\rho_{B|i}$ is a local dephasing in subsystem $A$ \cite{chita}.

Discord quantifies the disturbance induced by local measurements to  multipartite  states \cite{discrev,OZ,HV}. Even separable mixed states can have non-zero discord, while all quantum correlations reduce to entanglement for pure states. Let ${\cal C}$ denote the set of zero discord states, described as classically correlated. They take the form $\sum_{k_{1...n}} p_{k_{1...n}}\ketbra{k_{1...n}}$, where $\{p_{k_{1...n}}\}\geq0$, $\sum_{k_{1...n}}p_{k_{1...n}}=1$ and $\{\ket{k_{1...n}}=\ket{k_1}\otimes...\otimes\ket{k_n}\}$ is an arbitrary product basis. A class of measures quantifies discord as the pseudo-distance to the set ${\cal C}$ : $D^\delta(\rho)=\min\limits_{\sigma\in {\cal C}}\delta(\rho,\sigma)$ \cite{discrev}. In particular, the relative entropy of discord is given by $D(\rho_{A_1...A_n})=\min\limits_{\Phi^i}D_{\{\Phi^i\}}(\rho_{A_1...A_n}), D_{\{\Phi^i\}}(\rho_{A_1...A_n})=S(\rho_{A_1...A_n}||\Phi^i(\rho_{A_1...A_n}))$, where $\Phi^i =\otimes_{j=1}^n\Phi_{A_j}^{i_j}$ is the dephasing, i.e. a projective measurement, in $\{\ket{i_{1\ldots n}}\}$ \cite{modi}.  An alternative measure is given by the global discord $\overline{D}(\rho_{A_1...A_n})=\min\limits_{\{\Phi^i\}}\overline{D}_{\{\Phi^i\}}(\rho_{A_1...A_n}),\ \overline{D}_{\{\Phi^i\}}(\rho_{A_1...A_n}) = S(\rho_{A_1...A_n}||\Phi^i(\rho_{A_1...A_n}))-\sum_k S(\rho_{A_k}||\Phi^i_{A_k}(\rho_{A_k}))$, where $\Phi^i_{A_k}(\rho_{A_k})=\sum_i |i_k\rangle\langle i_k|\rho_{A_k} |i_k\rangle\langle i_k|$ \cite{global}.  These two quantities evaluate the disturbance induced by applying local dephasing to all the subsystems. Yet, disturbance also occurs when performing local measurements only on a fraction of the subsystems. This is captured by the asymmetric discord measures. For  bipartite states $\rho_{AB}$, by dephasing only on $A$, one has $\overline{D}_{B|A}(\rho_{AB})=\min\limits_{\{\Phi^i_A\}}\overline{D}_{\{\Phi^i_A\}}(\rho_{AB}), \overline{D}_{\{\Phi^i_A\}}(\rho_{AB}) = S(\rho_{AB}||\Phi^i_A(\rho_{AB}))-S(\rho_{A}||\Phi^i_A(\rho_{A}))$, which is equivalent to the original definition of discord \cite{OZ,HV}. We note that discord is alternatively defined as the minimum disturbance induced by generalized measurements (POVM), thus taking smaller values than the above introduced quantities \cite{discrev}. As this choice does not affect our results, we stick to the minimization over dephasing. From now on, the dephasing for all basis-dependent discord measures, e.g., $\Phi^i$ for $\overline{D}_{\{\Phi^i\}}(\rho_{A_1...A_n})$, is with respect to the reference basis under consideration.

We are now ready to  link  coherence and discord. It is evident from the above definitions that $D^{\delta}(\rho_{A_1...A_n})\leq C^{\delta}(\rho_{A_1...A_n})$ (for the relative entropy, see \cite{pradisc}) for any choice of the reference basis. In fact, this measure of discord is the minimum amount of coherence in any product basis \cite{discrev}. Let us then consider a setting where a coherent state $\rho_A$ is coupled to an initially uncorrelated incoherent ancilla $\tau_B$. If the coupling is an incoherent operation, then there is a bound on the amount of discord generated:

{\it Result 1 --} {\it For any contractive pseudo-distance $\delta$  the amount of discord created between a state $\rho_A$ and an incoherent ancilla $\tau_B$ by an incoherent operation $\Lambda_{\textsc{IC}}$ is upper bounded by the coherence of $\rho_A$:
\begin{equation}\label{res1}
D^\delta(\Lambda_{\mathrm{IC}}(\rho_A\otimes\tau_B))\leq C^\delta(\rho_A).
\end{equation}
}\\
{\it Proof -- } Suppose $\tau_A$ is the closest incoherent state to $\rho_A$. By exploiting the contractivity of $\delta$ and that  $\Lambda_{\mathrm{IC}} (\tau_A\otimes\tau_B)\in\mathcal{C}$, we have $C^{\delta} (\rho_A)=\delta(\rho_A, \tau_A)=\delta(\rho_A\otimes\tau_B, \tau_A\otimes\tau_B)\geq\delta(\Lambda_{\mathrm{IC}}(\rho_A\otimes\tau_B), \Lambda_{\mathrm{IC}}(\tau_A\otimes\tau_B))\geq D^{\delta}(\Lambda_{\mathrm{IC}}(\rho_A\otimes\tau_B))$. Q.E.D.

As $d_B\geq d_A$, the bound can be saturated for $\delta$ being the quantum relative entropy and the Bures distance. Here is the proof. Let us now fix $d_B\geq d_A$, $\delta$ being the quantum relative entropy, and $\tau_B=\ketbra{i_0}$. By results in \cite{ger2}, there exists an incoherent unitary gate (a generalized controlled-NOT gate) such that $E^{\delta}(\Lambda_{\mathrm{IC}}(\rho_A\otimes\tau_B))=C^{\delta}(\rho_A)$, where $E^{\delta}$ is the relative entropy of entanglement. Since $D^{\delta}(\rho)\geq E^{\delta}(\rho)$, we obtain $D^{\delta}(\Lambda_{\mathrm{IC}}(\rho_A\otimes\tau_B))\geq C^{\delta}(\rho_A)$. Thus,
$D^{\delta}(\Lambda_{\mathrm{IC}}(\rho_A\otimes\tau_B))=C^{\delta}(\rho_A)$. If $\tau_B$ is a different  incoherent state, we just perform the incoherent operation that replaces it with $\ketbra{i_0}$, and the proof follows. A similar argument holds for $\delta$ the Bures distance.

It is already known that  coherence can be converted to entanglement\cite{ger2}. Since discord is a quantum correlation beyond entanglement, a natural question arises: is it possible to transform coherence to discord without generating entanglement? We answer this question by providing such an example. Let us start with the state $\ketbra{+}\otimes\ketbra{0}$, where $\ket{+}=\frac{1}{\sqrt{2}}\ket{0}+\frac{1}{\sqrt{2}}\ket{1}$. Then we perform the incoherent operation $\varepsilon(\rho)=pU_{\mathrm{CX}}\rho U_{\mathrm{CX}}^{\dagger}+(1-p)\frac{\mathbb{I}}{4}$, $U_{\mathrm{CX}}$ is the controlled-NOT gate $U_{\mathrm{CX}}(\ket{i}\otimes\ket{j})=\ket{i}\otimes\ket{i\oplus j}$. The resulting state is a Werner state $p\ketbra{\Phi^+}+(1-p)\frac{\mathbb{I}}{4}$, with $\ket{\Phi^+}=\frac{1}{\sqrt{2}}\ket{00}+\frac{1}{\sqrt{2}}\ket{11}$. By tuning $0<p\leq\frac{1}{3}$, one can generate discord without any entanglement. It is easy to check that this case respects Result 1. An open question is whether or not the bound can be saturated for some separable output state. We leave this issue for future research.

Result 1 shows that the initial coherence of the global state (here equivalent to the coherence of subsystem $A$) upper bounds the creation of discord by incoherent operations. It is possible to extend the link to multipartite scenarios and tighten the bound by focusing on the relative entropy of coherence. Given an arbitrary channel $\Lambda(\cdot)$, we define the consumption of quantity $X$ in system $Y$ under $\Lambda(\cdot)$ as $\Delta X(\rho_Y)= X(\rho_{Y})-X(\Lambda(\rho_{Y}))$. When $\Delta X(\rho_Y)$ is negative, $-\Delta X(\rho_Y)$ can be understood as the production of $X$. For example, $\Delta C(\rho_A)$ denotes the coherence consumption of system $A$. We note that this quantity is closely related to the decohering power of a channel introduced by Mani and Karimipour~\cite{MK}, which corresponds to the maximum achievable coherence consumption when the input $\rho_A$ is optimized over the set of maximally coherent states. Also observe that $\overline{D}_{\{\Phi^i\}}(\rho_{A_1...A_n})=C(\rho_{A_1...A_n})-\sum_k C(\rho_{A_k})\geq 0$, we find $\Delta C(\rho_{A_1...A_n})=\sum_k\Delta C(\rho_{A_k})+\Delta \overline{D}_{\{\Phi^i\}}(\rho_{A_1\ldots A_n})$. If the channel is a global incoherent operation, coherence is consumed, $\Delta C(\rho_{A_1\ldots A_n})\geq0$. Thus:

{\it Result 2 -- Given an incoherent operation $\Lambda_{\mathrm{IC}}$ applied to an uncorrelated multipartite system $\rho_{A_1...A_n}=\otimes_{i=1}^n\rho_{A_i}$, the production of global discord is upper bounded by the coherence consumption of the subsystems:
\begin{equation}\label{res2}
\overline{D}(\Lambda_{\mathrm{IC}}({\rho}_{A_1...A_n}))\leq\sum_k\Delta C(\rho_{A_k}).
\end{equation}
}
{\it Proof --} Since $\sum_k\Delta C(\rho_{A_k})+\Delta \overline{D}_{\{\Phi^i\}}(\rho_{A_1\ldots A_n})\geq 0$, one has $\sum_k\Delta C(\rho_{A_k})\geq \overline{D}_{\{\Phi^i\}}(\Lambda_{\mathrm{IC}}({\rho}_{A_1...A_n}))-\overline{D}_{\{\Phi^i\}}(\rho_{A_1...A_n})
=\overline{D}_{\{\Phi^i\}}(\Lambda_{\mathrm{IC}}({\rho}_{A_1...A_n}))
\geq \overline{D}(\Lambda_{\mathrm{IC}}({\rho}_{A_1...A_n}))$. Q.E.D.

The bound is saturated if the least disturbing dephasing in $\overline{D}(\Lambda_{\mathrm{IC}}({\rho}_{A_1...A_n}))$ is the one on the reference basis, and $\Lambda_{\text{IC}}$ is a unitary (thus reversible) operation.

Following the same line of reasoning, the creation of asymmetric discord is bounded by the coherence consumption of the subsystems:\\
{\it Result 3 - Given an A-incoherent operation $\Lambda_{\mathrm{IC}}^{B|A}$ applied to an uncorrelated bipartite state $\rho_{AB}=\rho_A\otimes\rho_B$, the production of asymmetric discord is upper bounded by the coherence consumption in $A$,
\begin{equation}\label{res3}
\overline{D}_{B|A}(\Lambda_{\mathrm{IC}}^{B|A}(\rho_{AB}))\leq\Delta C(\rho_{A}).
\end{equation}
}
{\it Proof --} We note that $\overline{D}_{\{\Phi^i_A\}}(\rho_{AB}) =C_{B|A}(\rho_{AB})-C(\rho_A)$, thus $\Delta C_{B|A}(\rho_{AB})-\Delta C(\rho_A)=\Delta \overline{D}_{\{\Phi^i_A\}}(\rho_{AB})$. The monotonicity of $C_{B|A}$ under $\Lambda^{B|A}_{\mathrm{IC}}$ implies $\Delta C_{B|A}(\rho_{AB}) \geq 0$. As $\overline{D}_{\{\Phi^i_A\}}(\rho_{AB})=0$, one has $\sum_k\Delta C(\rho_{A_k})\geq \overline{D}_{\{\Phi^i_A\}}(\Lambda_{\mathrm{IC}}^{B|A}(\rho_{AB}))-\overline{D}_{\{\Phi^i_A\}}(\rho_{AB})
=\overline{D}_{\{\Phi^i_A\}}(\Lambda_{\mathrm{IC}}^{B|A}(\rho_{AB}))
\geq \overline{D}_{B|A}(\Lambda_{\mathrm{IC}}^{B|A}(\rho_{AB}))$. Q.E.D.


We observe that defining discord by minimizing over POVM would still lead to the same upper bounds in Results 1-3. It is also notable that for pure states the results provide a link between coherence consumption and creation of entanglement. In the following, we discuss two quantum information protocols to show how the interplay between coherence and discord works.

\begin{figure}
\begin{center}
\includegraphics[scale=0.25]{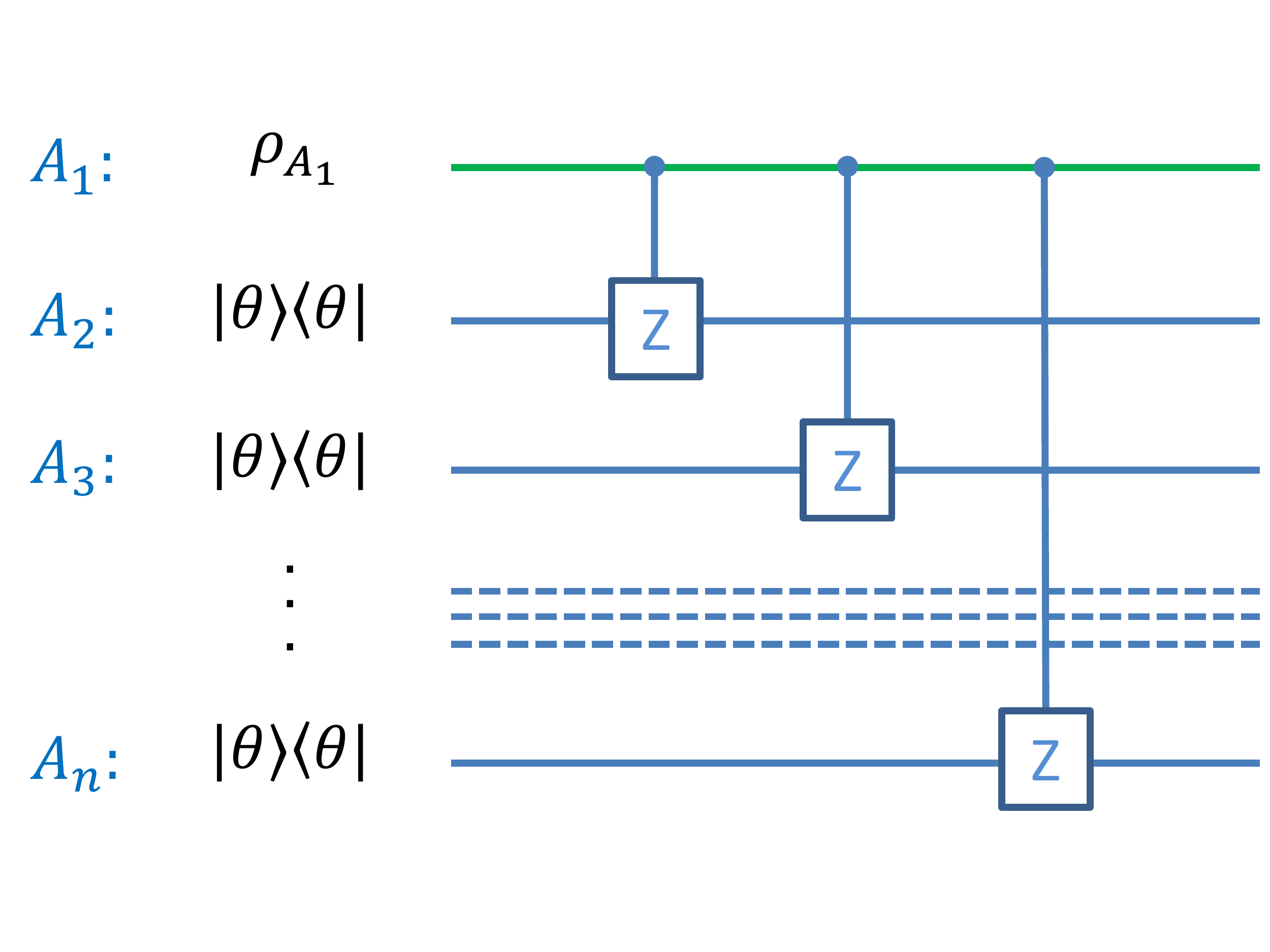}
\end{center}
\vspace{-20pt}
\caption{Preparation of a quantum correlated state. A sequence of controlled-$Z$ gates  is applied to generate quantum correlations in a quantum system that is initialized in a product state. The sequence represents a global incoherent operation for the register, i.e., the coherence of the global state is non-increasing (it is constant in such a case). Each controlled gate consumes the coherence of the control and target qubits and increases the global discord of the state.  \label{fig:fig1}}
\end{figure}

\textit{Preparation of a quantum correlated state --} Let us consider a register of $n$ qubits initialized in the product state $\rho_{A_1...A_n}=(p\frac{\mathbb{I}}{2}+(1-p)\ketbra{\theta})\otimes\ketbra{\theta}^{\otimes n-1}, \ket{\theta}=\cos\theta\ket{0}+\sin\theta\ket{1}, p\in[0 ,1],\theta\in[0, \pi]$, where the noise in the first qubit is added to distinguish discord from entanglement. By applying the network in Fig. \ref{fig:fig1}, we aim to prepare a quantum correlated state, which can be then employed to run a quantum computation. In particular, for $p=0,\, \theta=\pi/4$, the scheme generates a maximally entangled (graph) state with applications in quantum computation \cite{oneway,nielsen,bell} and metrology \cite{rosenkranz}. The state preparation consists of a sequence of two-qubit controlled-$Z$  gates $U_{\mathrm{CZ}}(\ket{i}\otimes\ket{j})=(-1)^{ij}\ket{i}\otimes\ket{j}$ which create quantum correlations between the register qubits. This is an incoherent operation in the standard computational basis $|i\rangle\otimes |j\rangle$.

\begin{figure}[!htb]
\begin{center}
\includegraphics[scale=0.9]{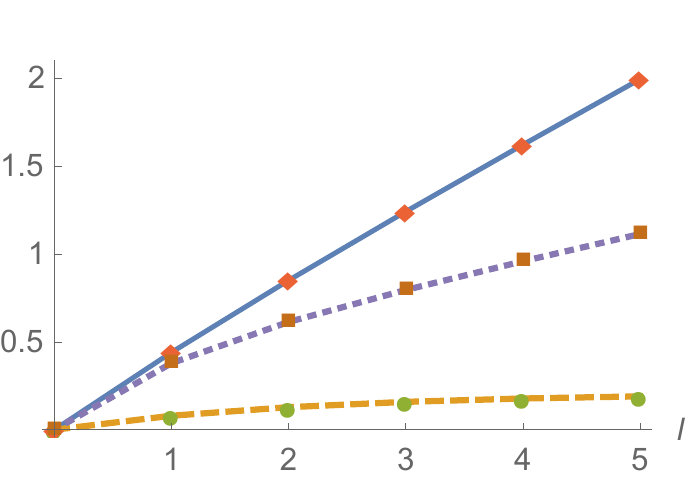}
\end{center}
\caption{Coherence consumption bounds discord creation. We plot the values of the total coherence consumption of the subsystems $l\times \Delta C(\rho_{A_k})+\Delta C^l(\rho_{A_1})$ (solid linked diamonds), the created global discord $\overline{D}(\rho^l_{A_1...A_n})$ (dotted linked squares) obtained by numerical optimization, and the coherence consumption (dashed linked discs) on the control qubit $\Delta C^l(\rho_{A_1})$  after $l$ controlled-$Z$ gates for $p=0.2$ and $\theta=0.45$. \label{fig:plot3}}
\end{figure}
Let $\rho^l_{A_1...A_n}$ denote the state after performing the $l$-th controlled-$Z$. From Eq. \ref{res2}, one has (full derivation in supplementary material)
\begin{equation}\label{eq5}
\overline{D}(\rho^l_{A_1...A_n})\leq l\times \Delta C(\rho_{A_k})+\Delta C^l(\rho_{A_1}),
\end{equation}
where $\Delta C^l(\rho_{A_1})=C(\rho_{A_1})-C(\rho^l_{A_1})$ denotes the total coherence consumption on the control qubit after applying the $l$-th controlled-$Z$ gate and $\Delta C(\rho_{A_k}), k=2,...,l+1$ is the coherence consumption on each target qubit for each of the $l$ applications of the controlled-$Z$ . We calculate the values of these quantities and plot them in Fig. \ref{fig:plot3}, which highlights the expected relation between coherence consumption and global discord production. Due to the symmetry of the problem, each controlled-$Z$ consumes an equal share of coherence in each target, i.e., $\Delta C(\rho_{A_k})$ is the same for each $k\geq2$.

\begin{figure}[!htb]
\begin{center}
\includegraphics[scale=0.3]{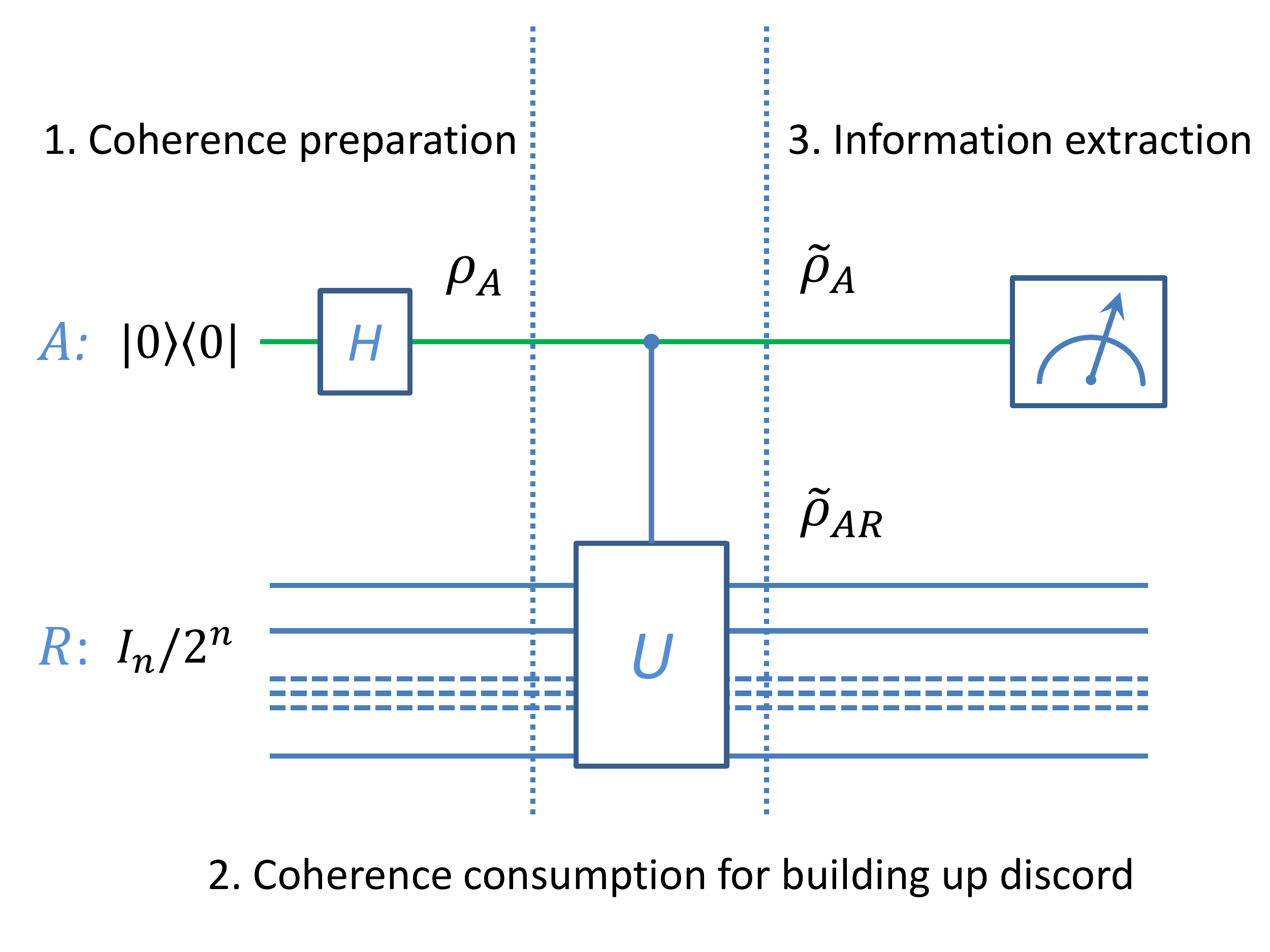}
\end{center}
\caption{DQC1 model. The algorithm consists of: (1) preparing   coherence in an ancillary pure qubit; (2) applying a controlled operation, which in general creates discord between the ancilla and a maximally mixed register by consuming the coherence of the ancilla; (3) obtaining information about $\text{Tr}[U]$ by a polarisation measurement in the ancilla: $\langle\sigma_x+ i \sigma_y\rangle_{\widetilde{\rho}_A} = \text{Tr}[U]/2^n$.\label{fig:DQC1}}
\end{figure}

\textit{Coherence consumption and discord production in DQC1 -- }This model is a quantum computation which estimates the normalized trace of a unitary matrix $U$ with exponential speed-up with respect to the known classical algorithms \cite{dqc1}. It employs highly mixed qubits such that vanishing entanglement is generated in the multipartite state. The protocol is presented in Fig. \ref{fig:DQC1}. The algorithm requires an $n+1$ qubit system split into one ancilla $\ket{0}\bra{0}_A$ (in fact, the result applies to any $\rho_A$) and an $n$-qubit maximally mixed register, $\rho_R=\mathbb{I}/2^n$, initially in a product state. This scheme consists of first applying a Hadamard gate on the register, followed by a controlled-U gate that correlates this register with the ancilla. Subsequent measurements of the ancilla in appropriate basis enables estimation of the normalized trace of $U$: $\langle\sigma_x+ i \sigma_y\rangle_{\widetilde{\rho}_A} =\text{Tr}[U]/2^n$.

It is conjectured that discord in the state just prior to measurement, $\widetilde{\rho}_{AR}$, is the resource for the protocol \cite{datta,datta1,eastin,dakic}. Here we provide a further viewpoint by studying DQC1 in terms of the interplay between coherence and discord.

We fix the reference basis to be the computational basis for the ancilla and the eigenbasis of $U$ for the register. Our reference basis choice makes the controlled-$U$ an incoherent operation. Also, the state of the register is always maximally mixed, thus remains incoherent throughout the algorithm. The protocol requires coherence to be built in the ancilla state, $C(\rho_A)=1$, then to be consumed in order to correlate the ancilla and the register.

From Eq.\eqref{res2}, we have  $\Delta C(\rho_A)+\Delta C(\rho_R)\geq -\Delta \overline{D}(\rho_{AR})$. By noting that $\Delta C(\rho_R)=0$ and  $\overline{D}(\rho_{AR})=0$, we find that the coherence consumption in the ancilla bounds the generated global discord:
\begin{equation}
 \overline{D}(\widetilde{\rho}_{AR})\leq\Delta C(\rho_A).
\end{equation}
Let us focus on the asymmetric discord.  It is straightforward to see that controlled-$U$ is an A-incoherent operation, i.e. it maps the set $\mathcal{I}_{R|A}$ to itself. Since the global state before applying the controlled-$U$ is a product state, Eq. \ref{res3} implies
\begin{eqnarray}\label{zurek}
\overline{D}_{R|A}(\widetilde{\rho}_{AR})\leq\Delta C(\rho_A).
\end{eqnarray}

Previous claims about discord as a resource in DQC1 have several flaws. One is that as $U=e^{i\phi}M, M^2=\mathbb{I}$, there is no discord in $\widetilde{\rho}_{AR}$ \cite{dakic}. Nevertheless, no classically efficient method of estimating the normalized trace of such $U$ is known. We note coherence can generally be consumed in this case. In fact, $\Delta C(\rho_{A})=H_2(\frac{1-|\text{Tr}[U]|/2^n}{2})$, where $H_2(x) = -x \log(x) - (1-x) \log(1-x)$ is the binary Shannon entropy. Therefore no coherence is consumed if and only if $U = e^{i\phi}\mathbb{I}$ for some $\phi$. This motivates an interesting question of whether a classical algorithm can evaluate the normalized trace of such $U$. The question, however, is surprisingly non-trivial. This is because the input of the DQC1 algorithm is not a matrix representation of $U$, but rather some classical description for a polynomial sized quantum circuit (sequence of one and two-qubit gates) that generates $U$~\cite{datta3}. If true, it would indicate that coherence consumption is non-zero in DQC1 if and only if quantum processing has a computational advantage.


\textit{Conclusion} --  In this work, we investigated the interplay between coherence and discord. We proved several results bounding the amount of discord which can be created {\it ex nihilo} by incoherent operations in a multipartite system. We showed that coherence of the subsystems must be consumed in order to create genuinely quantum correlations between them. Establishing the cost of creating quantum resources is a necessary requirement for quantum information processing. Here we proved that in a scenario where coherence of the global state is not a freely available resource, which means that only incoherent operations are allowed, discord is created only if coherence of the subsystems is consumed.

This also motivates a line of future research. Do there exists situations where quantum correlations be considered a natural resource for generating states of high coherence? Together these relations could have particular importance in the study of open systems, where systems under study constantly interact with the environment. 

\section*{Acknowledgments}
We thank Tillman Baumgratz, Andrew Garner, Jayne Thompson, Mark Wilde and Andreas Winter for fruitful discussions; and gratefully acknowledge National Research Foundation (NRF), NRF-Fellowship (Reference No: NRF-NRFF2016-02), the EPSRC (UK) Grant EP/L01405X/1, the John Templeton Foundation  Grant 54914, the National Research Foundation, the Ministry Education in Singapore Grant and the Academic Research Fund Tier 3 MOE2012-T3-1-009, the National Basic Research Program of China Grant 2011CBA00300, 2011CBA00302 and the National Natural Science Foundation of China Grant 11450110058, 61033001, 61361136003, the Leverhulme Trust and the Oxford Martin School and the Wolfson College, University of Oxford for funding this research.

\clearpage

 \appendix*

\section{Supplemental Material}
\begin{center}
{\large{\bf {Converting Coherence to Quantum Correlations}}}

\quad \\

{\normalsize Jiajun Ma, Benjamin Yadin, Davide Girolami, Vlatko Vedral, and Mile Gu}

\end{center}

\setcounter{equation}{0}
\subsection{Derivation of Eq.4 in the paper}
In the current scenario, the system is initialized to be $\rho_{A_1...A_n}=(p\frac{\mathbb{I}}{2}+(1-p)\ketbra{\theta})\otimes\ketbra{\theta}^{\otimes n-1}$, where $\ket{\theta}=\cos\theta\ket{0}+\sin\theta\ket{1}$ and $\theta\in[0, \pi], p\in[0,1]$. Let $\rho^l_{A_1...A_n}$ denote the state after applying $l$-th controlled-$Z$ gate to the state. Clearly, $\rho^l_{A_1...A_n}=\rho^l_{A_1...A_{l+1}}\otimes\ketbra{\theta}^{\otimes n-l-1}$, where $\rho^l_{A_1...A_{l+1}}=\tr{}_{A_{l+1}...A_n}[\rho^l_{A_1...A_n}]$. Meanwhile, we have
\begin{equation}
\rho^l_{A_1...A_{l+1}}=p\times\sigma^l_{A_1...A_{l+1}}+(1-p)\times\tau^l_{A_1...A_{l+1}},
\end{equation}
where
\begin{equation}
\sigma^l_{A_1...A_{l+1}}=\frac{1}{2}\ketbra{0}\otimes\ketbra{\theta}^{\otimes l}+\frac{1}{2}\ketbra{1}\otimes\ketbra{-\theta}^{\otimes l},
\end{equation}
and
\begin{equation}
\tau^l_{A_1...A_{l+1}}=\ket{\psi}^l\bra{\psi}^l,
\end{equation}
with $\ket{\psi}^l=\cos\theta\ket{0}\ket{\theta}^{\otimes l}+\sin\theta\ket{1}\ket{-\theta}^{\otimes l}$. Thus the reduced state of $A_1$ is given by
\begin{equation}
\rho^l_{A_1}={\footnotesize
\begin{pmatrix}
\frac{p}{2}+(1-p)\cos^2\theta & (1-p)\sin\theta\cos\theta\sclp{-\theta}{\theta}^l\\
(1-p)\sin\theta\cos\theta\sclp{\theta}{-\theta}^l & \frac{p}{2}+(1-p)\sin^2\theta
\end{pmatrix}\nonumber}
\end{equation}
Since $|\sclp{-\theta}{\theta}|\leq 1$, the modulus of the off-diagonal elements of $\rho^l_{A_1}$  decreases with $l$. This means that the coherence consumption of $A_1$, denoted by $\Delta C^l(\rho_{A_1})$, keeps increasing as controlled-$Z$ gates are applied. For the other qubits labelled $k=2,3,...,l+1$, the marginal state  is given by $\rho^l_{A_k}=\cos^2\theta\ketbra{\theta}+\sin^2\theta\ketbra{\theta}$. Hence, the total coherence consumption of these qubits is $l\times\Delta C(\rho_{A_i})$.
Since controlled-$Z$ is an  incoherent operation, by Result 2 of the main text, the total accumulated coherence consumption after  the $l$-th controlled-$Z$ gate is given by
\begin{equation}
\overline{D}_{\{\Phi^i\}}(\rho^l_{A_1...A_n})=\Delta C^l(\rho_{A_1})+l\times\Delta C(\rho_{A_k}),
\end{equation}
as the initial state has no correlation, $\Delta \overline{D}(\rho_{A_1...A_n})=-\overline{D}(\rho^l_{A_1...A_n})$. Finally, since $\overline{D}(\rho^l_{A_1...A_n})\leq \overline{D}_{\{\Phi^i\}}(\rho^l_{A_1...A_n})$, one has
\begin{equation}
\overline{D}(\rho^l_{A_1...A_n})\leq \Delta C^l(\rho_{A_1})+l\times\Delta C(\rho_{A_i}).
\end{equation}


\begin{thebibliography}{99}
\bibitem{note}The frameworks proposed in Refs.\cite{plenio,gour} differ in the definition of incoherent operations. We here embrace the viewpoint of Ref.\cite{plenio}.
\bibitem{QI}These states have been called quantum-incoherent states in \cite{chita}.
\end{thebibliography}
\end{document}